\documentclass[pdflatex,sn-mathphys]{sn-jnl}

\jyear{2023}

\begin{document} 
\title[Quantum error mitigation in the regime of high noise ]{Quantum error mitigation in the regime of high noise using deep neural network: Trotterized dynamics}

\author*[1]{\fnm{Andrey} \sur{Zhukov}}\email{zugazoid@gmail.com}
\author[1,2,3]{\fnm{Walter} \sur{Pogosov}}\email{walter.pogosov@gmail.com}

\affil[1]{\orgname{Dukhov Research Institute of Automatics (VNIIA)}, \orgaddress{\city{Moscow}, \postcode{127030}, \country{Russia}}}
\affil[2]{\orgname{Advanced Mesoscience and Nanotechnology Centre, Moscow Institute of Physics and Technology (MIPT)}, \orgaddress{\city{Dolgoprudny}, \postcode{141700}, \country{Russia}}}
\affil[3]{\orgdiv{Institute for Theoretical and Applied Electrodynamics}, \orgname{Russian Academy of Sciences}, \orgaddress{\city{Moscow}, \postcode{125412}, \country{Russia}}}

\abstract{
We address a learning-based quantum error mitigation method, which utilizes deep neural network applied at the postprocessing stage, and study its performance in presence of different types of quantum noises. We concentrate on the simulation of Trotterized dynamics of 2D spin lattice in the regime of high noise, when expectation values of bounded traceless observables are strongly suppressed. By using numerical simulations, we demonstrate a dramatic improvement of data quality for both local weight-1 and weight-2 observables for the depolarizing and inhomogeneous Pauli channels. At the same time, the effect of coherent $ZZ$ crosstalks is not mitigated, so that in practise crosstalks should be at first converted into incoherent errors by randomized compiling. 
}

\keywords{quantum algorithms, machine learning, neural networks, quantum error mitigation, Ising model, NISQ processors, Trotter evolution}

\maketitle

\section{Introduction}

Current quantum hardware is characterized by the noise rate which makes it not yet possible to implement fault-tolerant quantum computing. In addition, available number of qubits in quantum devices is not large enough to achieve this goal. Such quantum computers belong to the class of noisy intermediate-scale  quantum (NISQ) machines, whose performance is limited \cite{preskill2018quantum}. Nevertheless, recently the utility of quantum computing before fault tolerance was argued to be demonstrated with superconducting quantum computer using quantum error mitigation technique \cite{kim2023evidence}. In general, various quantum error mitigation schemes have been developed in order to improve the performance of NISQ devices without error correction, see, e.g., Refs. \cite{
li2017efficient, Bravyi, endo2018practical, kandala2019error, he2020zero, vovrosh2021simple, urbanek2021mitigating, strikis2021learning, czarnik2021error, sun2021mitigating, cai2021multi, mari2021extending, cai2022quantum, perrin2023mitigating, kim2023evidence, kim2023scalable, bultrini2023unifying, van2023probabilistic}. The most prominent examples are probabilistic error cancellation \cite{Bravyi}, which requires a precise knowledge of error model, and zero noise extrapolation \cite{li2017efficient,Bravyi}, which is more error-agnostic. It is known that quantum errors can be roughly divided into two classes -- incoherent and coherent errors. Incoherent errors are due to the decoherence and imperfect gate executions, while coherent errors are mainly due to the miscalibration in control parameters as well as the residual coupling between qubits. Coherent errors being systematic conserve quantum state purity, while this is not the case for incoherent errors. By using randomized compiling (twirling) it is possible to transform coherent errors into incoherent ones, which are easier to handle \cite{kern2005quantum,wallman2016noise,hashim2020randomized,perrin2023mitigating}. 

Despite of the impressive progress in error mitigation strategies, most of widely adopted practical methods, in general, are not so efficient in the regimes of high noise, when expectation values of bounded traceless observables tend to cluster around zero. In particular, popular zero noise extrapolation is based on collecting data at different noise levels (noise parameters), which can be increased artificially beyond the base level by stretching pulse duration used for quantum gates or by inserting noisy identity gates into the circuit, and extrapolating the result to the zero noise level. Such an approach faces a problem of uncertainty of the extrapolation in the regimes of high noise, when the circuit fault rate is large \cite{czarnik2021error,bultrini2023unifying}. 

We have recently proposed a learning-based method of quantum error suppression, which uses deep neural network (DNN) at the postprocessing stage \cite{zhukov2022quantum}. The approach can be applied to quantum circuits with repetitive gate layers, such as circuits for Trotterized quantum simulation. The Trotterized time evolution is an important circuit benchmark, which was very recently suggested to demonstrate the utility of quantum computing with NISQ devices \cite{kim2023evidence}. Our basic idea is to train a quantum network for smaller number of Trotter steps $N_1$ compared to the the same number $N_2 > N_1$ in the target circuit of interest, i.e., training is performed for shallower circuit, for which the output is known. The training data can be obtained either from the same quantum computer, since in this case they will be less noisy compared to the data obtained from the target circuit ($N_1 < N_2$). Note that the data quality can be additionally improved by a direct zero noise extrapolation. Another possibility is to rely on classical simulation methods, such as tensor network approaches to produce data for as much Trotter steps as possible. For the training circuit, $N_1$ Trotter steps for total evolution time $T$ are supplemented by noisy identity gates, each gate being represented by $N_2-N_1$ "empty" Trotter layers, from which all rotations are removed, while all CNOTs are left. Noise rate as well as the circuit structure in the case of such a combined circuit mimic those for the target circuit of interest consisting of $N_2$ Trotter steps for the same total evolution time $T$. Training circuits are constructed for different initial conditions of the simulated quantum model, which can be a spin model or a fermionic model. In Ref. \cite{zhukov2022quantum} we performed proof-of-principle experiments using superconducting quantum computers of IBM Quantum Experience for both the transverse-field Ising and XY models in 1D and observed a significant error reduction. The advantage of our method is that there is no need to boost the noise for deep circuits, for which observables are already strongly suppressed. Note that our protocol can be treated as a combination of a learning-based technique \cite{strikis2021learning,czarnik2021error} and the method utilizing a noise-estimation circuits \cite{urbanek2021mitigating}, while a new ingredient of our approach is a machine learning algorithm in the context of Trotterized evolution. In general, machine learning technology was found to be useful for such tasks as the verification of quantum devices \cite{lennon2019efficiently}, quantum error correction \cite{nautrup2019optimizing,baireuther2018machine, andreasson2019quantum}, quantum control \cite{kalantre2019machine, bukov2018reinforcement,niu2019universal}, quantum state classification \cite{babukhin2019nondestructive, carrasquilla2017machine}, and quantum state tomography \cite{altepeter2005photonic, torlai2018neural,neugebauer2020neural, lohani2020machine,sehayek2019learnability,palmieri2020experimental}.

The aim of the present article is to study the performance of the approach proposed in Ref. \cite{zhukov2022quantum} for different quantum noises. Performing numerical simulations instead of experiments with real hardware we can distinguish between different noises, which is of importance in the view of the fact that noise in superconducting quantum devices is known to have complex structure \cite{perrin2023mitigating,Hashim2023}. In particular, we examine depolarizing channel, inhomogeneous Pauli noise, as well as $ZZ$ crosstalks, the latter being especially harmful for fixed frequency superconducting qubits \cite{zhukov2019quantum,Babukhin2020,perrin2023mitigating}. As a benchmark, we consider 2D spin lattice described by the transverse-field Ising Hamiltonian. We observe a dramatic improvement of data quality by our method for all noise channels except of the crosstalk noise. We expect that the situation is similar for all coherent errors, including higher-weight crosstalk processes. For incoherent errors, the quality of improvement is limited essentially by the statistical uncertainties due to the probabilistic nature of measurements (shot noise), provided state preparation and measurements (SPAM) errors are neglected. This conclusion is made for both local weight-1 and weight-2 observables. We argue that, in practise, it is reasonable to combine our method with the randomized compiling in order to first transform coherent errors into an effective Pauli channel, see, e.g., Ref. \cite{Hashim2023}, and only after that to apply DNN to suppress the effect of this channel as well.

The paper is organized as follows. In Section II we formulate our model. In Section III we discuss the effect of depolarizing errors. In Section IV we consider the influence of non-depolarizing errors. We conclude in Section V.

\section{Model}

\subsection{Transverse-field Ising model}

As a test case, we consider the dynamics of 2D spin lattice described by the transverse-field Ising model Hamiltonian
\begin{equation}
    H =  -\sum_{j}h_{j}X_{j} -\sum_{<ij>}J_{ij}Z_{i}Z_{j},
    \label{eq:ham}
\end{equation}
where $h_{j}$ are local transverse fields and $J_{ij}$ are coupling constants which are nonzero only for nearest neighbors. In the present article we analyze the case of a square spin lattice and consider the disordered system, when all coupling constants are randomly distributed according to the Gaussian distribution. Hereafter we assume that $\bar{h}=2\bar{J}$ and standard deviations for $J$ and $h$ are $\bar{J}/2$ and $\bar{h}/2$, respectively.

The dynamics of the system starting from a given initial state can be simulated digitally using Trotter decomposition of the evolution operator. The total evolution time $T$ can be discretized into $N$ time steps $\delta t=T/N$. The evolution operator for each Trotter layer can be written in a standard way as a product of two operators given by
\begin{eqnarray}
 e^{-iH_{ZZ}\delta t}=\prod_{<ij>}R_{Z_i Z_j}(2J_{ij}\delta t),
 \\
 \nonumber
 e^{-iH_{X}\delta t}=\prod_{i}R_{X_i }(2h_{i}\delta t),
    \label{eq:ham2}
\end{eqnarray}
where $R_{Z_i Z_j}$ and $R_{X_i }$ are $ZZ$ and $X$ rotation gates, respectively. We are not concerned with the purely mathematical Trotterization error, but nevertheless we work in the regime, when both $\bar{J}\delta t$ and $\bar{h}\delta t$ are small. We also suppose that the spin lattice and the quantum processor share the same topology. Note that more complex Trotter decomposition schemes can be utilized to study the dynamics \cite{Ostmeyer2023} and even tensor-network approaches can be used for this purpose \cite{PhysRevResearch.5.023146}.

\begin{figure}
\begin{center}
\includegraphics[width=0.95\linewidth]{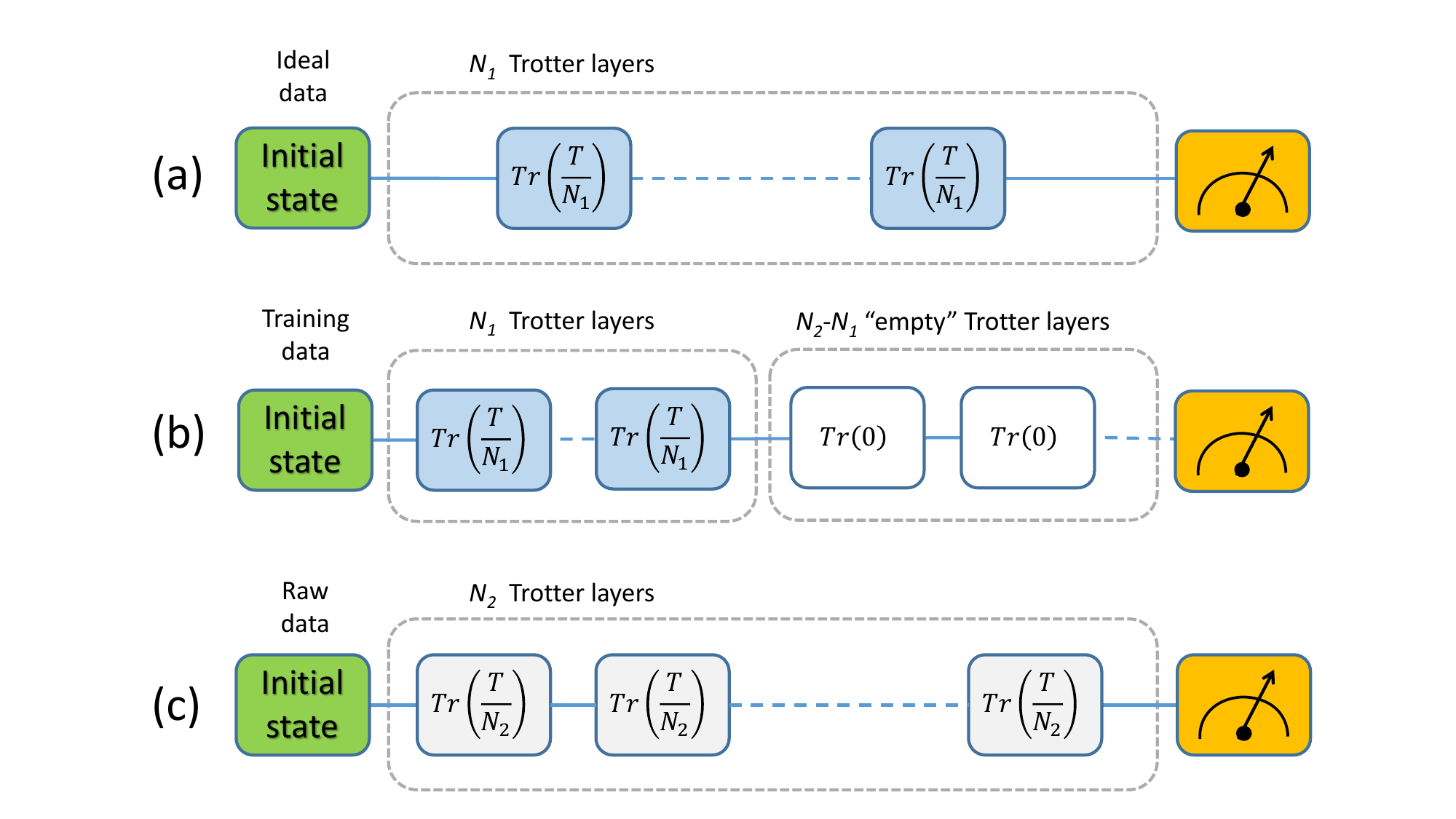}
\caption{Schematic view of the method we used. (a) - Generation of ideal or quasi-ideal data using classical computation or the same quantum computer corresponding to $N_1$ Trotter layers and different initial conditions. (b) - DNN training by adding $N_2-N_1$ "empty" Trotter layers to the quantum circuit and transforming such noisy data towards their ideal counterparts. (c) applying trained DNN to the noisy data corresponding to $N_2$ Trotter layers.}
\label{fig:scheme}
\end{center}
\end{figure}

\subsection{DNN training and structure}

Let us assume that the quantum circuit for $N_1$ Trotter steps with $\delta t_1=T/N_1$ is classically simulable using some known technique. This can be used for DNN training. In our studies, for illustrative purposes, we analyze the regime of a small or moderate number of qubits, so that we use a brute-force classical calculations for DNN training. Another possibility is to run the same quantum computer for $N_1$ Trotter steps and to utilize the output data for training. Their quality can be additionally improved using zero-noise extrapolation. At the same time, our target circuit we want to implement corresponds to $N_2 > N_1$ Trotter steps and $\delta t_2=T/N_2$, which cannot be simulated classically. At training stage we incorporate in our circuit extra identity gates constructed on the basis of Trotterized evolution (they are executed after the original $N_2$ Trotter layers), which leave the state nominally unchanged. We will consider $N_2-N_1$ "empty" Trotter steps, from which all rotations are removed, so that they contain only CNOTs, which are known to be the most noisy quantum gates. The idea is to make the training circuit structure as close to the target circuit structure as possible. DNN is trained to transform noisy data towards exact data. For such data we use either weight-1 or weight-2 observables. The quality criterion for DNN is the mean square error (MSE) between the individual expectation values of observables, such as spins magnetization. In order to obtain an informative training set, we consider different initial conditions of the spin lattice, which are eigenstates of the computational basis, and different time instances $t$ from the interval $(0,T)$. The whole idea is illustrated in Fig. \ref{fig:scheme}, where $Tr(\delta t)$ stands for the single Trotter layer with time step $\delta t$. Each training circuit corresponds to its own initial condition and time instance. Of course, this implies an exponential training overhead as the number of such initial states is exponentially large in terms of qubit number $n$. In order to reduce additional sampling costs (number of additional runs required), we use Monte-Carlo procedure and show that the scaling of our method is not exponential.

\begin{figure}
\begin{center}
\includegraphics[width=0.8\linewidth]{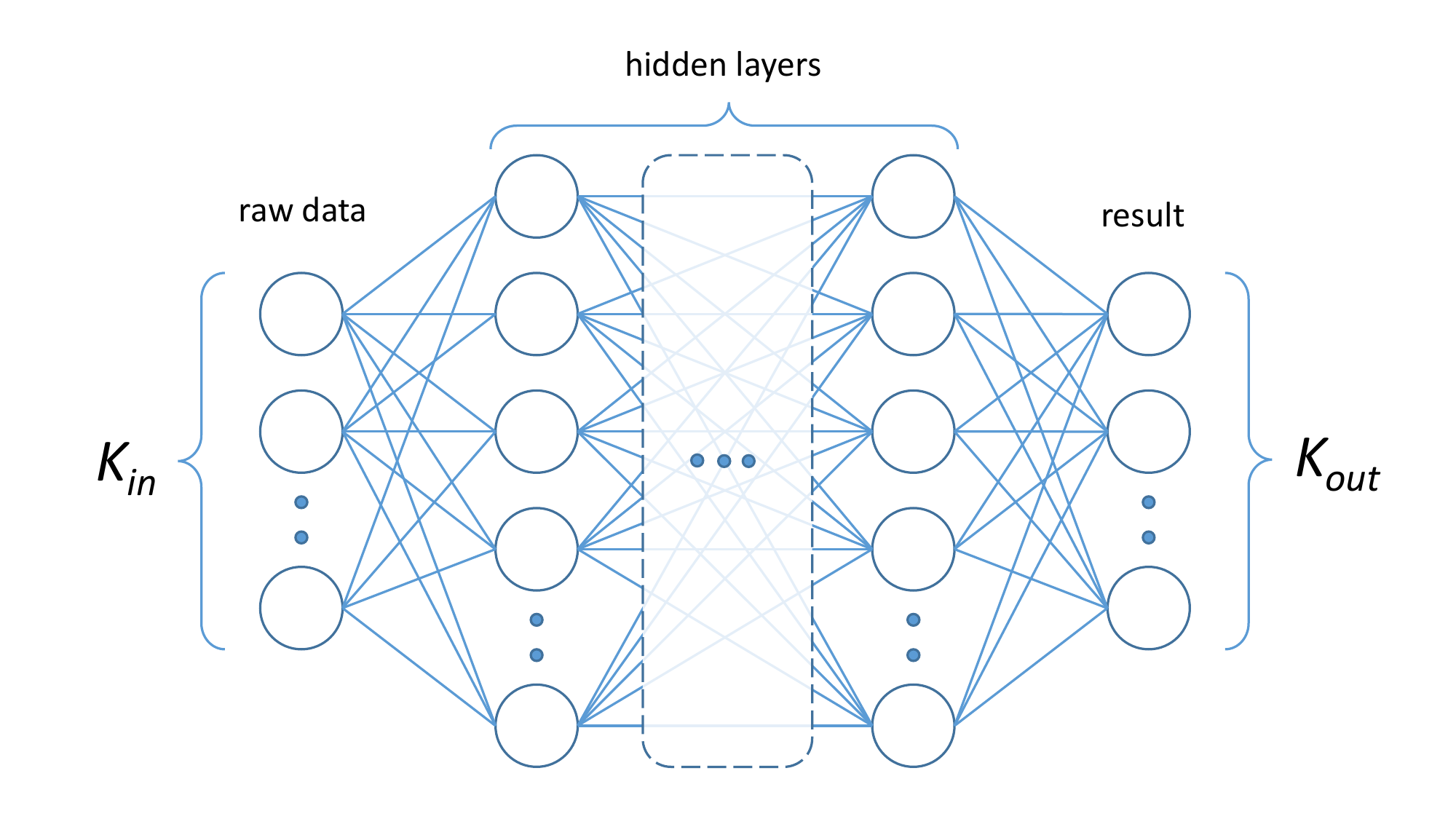}
\caption{The schematic view of the DNN structure. The number of input (output) neurons is $K_{in}$ ($K_{out}$). Several hidden layers can be used. In our illustrative simulations we used three hidden layers each consisting of 1000 neurons. The sigmoid activation function after both hidden
and output layers is utilized, while ReLu activation function is used for hidden layers. }
\label{fig:dnn}
\end{center}
\end{figure}

DNN structure is shown schematically in Fig. \ref{fig:dnn}. The number of input (output) neurons is equal to the number of input (output) quantities, such as individual spin magnetizations. It also has three hidden layers each containing 1000 neurons. The sigmoid activation function is used for both input and output layers, while ReLu activation function is used for hidden layers. Notice that it is also possible to utilize physics-inspired activation functions, which could further improve the results \cite{PRXQuantum.2.040355}, but this is beyond the scope of the present paper.  The DNN is trained to minimize the MSE between the exact data and data from the training set using Adam algorithm.

\section{Depolarizing errors}

In the present Section, we consider a depolarizing noise, which results in mixing of an initial qubit state with a completely depolarized state. This model is represented by the single-qubit quantum channel
\begin{equation}
    \Phi^{depol}_{1}(\rho,p_1) = (1 - p_1)\rho + \frac{p_1}{2}I,
\end{equation}
For two-qubit gates the quantum channel can be constructed out of single-qubit quantum channels in a following way 
\begin{equation}
    \Phi^{depol}_{2}(\rho,p_2) = (\Phi^{depol}_{1} (\rho,p_2) \otimes \Phi^{depol}_{1} (\rho,p_2))(\rho).
\end{equation}
This equation assumes uncorrelated errors on different qubits. For simplicity, we ignore SPAM errors. We use qiskit to simulate the quantum circuits.

\subsection{Weight-1 observables}

As a first example, we consider weight-1 observables, such as individual magnetizations of spins in $z$ direction, $\langle Z_j \rangle$. We perform simulations for the 2D square lattice containing $n=9$ spins. The feed-forward DNN has nine input neurons and nine output neurons corresponding to magnetizations of individual spins. For the training set, we consider magnetizations of spins for different initial conditions of the spin lattice and discrete instances of time from the interval $(0,T)$. We typically divide this interval into 300 equal segments. Monte Carlo sampling is used to circumvent the problem of exponentially large number of initial conditions. The sampling is performed among these initial conditions and randomly chosen time instances. Scaling issues are discussed in subsection C. In order to quantify the error mitigation efficacy, we evaluate the MSE between the raw data and the exact data, as well as the MSE between the improved data and the exact data. 

\begin{figure}
\begin{center}
\includegraphics[width=0.95\linewidth]{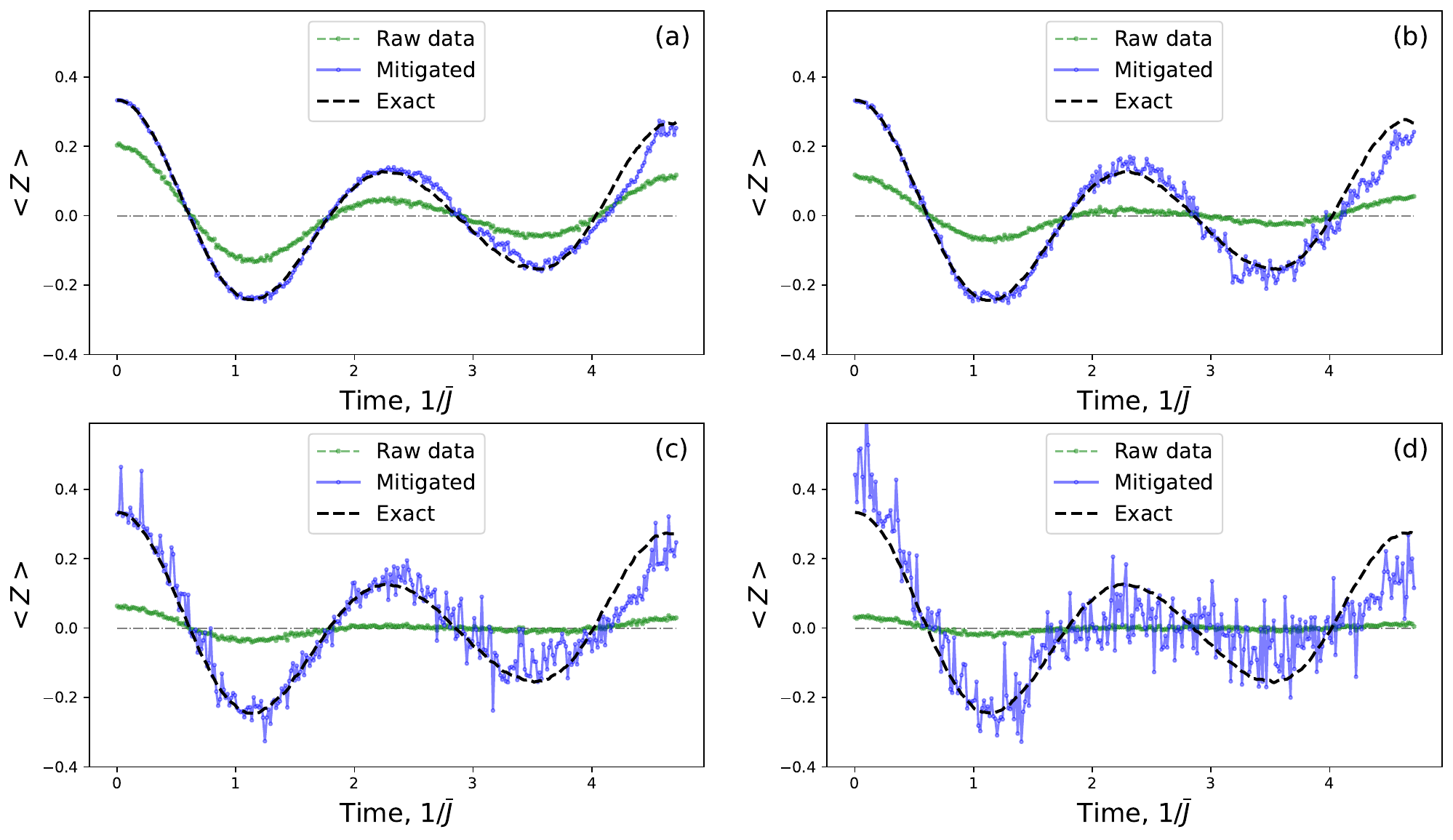}
    \caption{\label{fig:scheme_theory}
    The dependence of a mean magnetization in $z$ direction $\langle Z\rangle$ of 9-spin system on time for Trotter layer numbers $N_2=16$ (a), $N_2=32$ (b), $N_2=48$ (c), and $N_2=64$ (d) starting from the initial condition $\vert 000111000\rangle$ at $\bar{h}=2\bar{J}$. Trotter layer number at the training stage is $N_1=4$. Depolarizing error channel is assumed. }
\end{center}
\end{figure}

In Fig. \ref{fig:scheme_theory} we show the mean magnetization  $\langle Z \rangle = 1/n \sum_j \langle Z_j \rangle$ for a particular realization of disorder as a function of time for Trotter step numbers $N_2=16$ (a), $N_2=32$ (b), $N_2=48$ (c), and $N_2=64$ (d) for the initial condition $\vert 000111000\rangle$ (the qubits in the first and third rows are in the states 0, while the qubits in the second row are in the states 1). Note that this particular realization of disorder was used for all illustrative figures of this paper. Also note that in our illustrations, we choose different initial states and this is dictated by the fact that the dynamics should be rich enough to show interesting behaviour, since in many cases the dynamics of the quantities of interest turns out to be rather trivial. The training was performed for $N_1=4$ Trotter steps. Three curves are shown, which represent raw data corresponding to the error rates $p_1=10^{-4}$, $p_2=10^{-2}$, exact curve, and error-mitigated data. The error rates used in the simulations nearly correspond to the state-of-the-art superconducting quantum processors. The number of shots for each circuit is $N_{sh}=8192$. 

We see a dramatic improvement of the data quality by the DNN despite of the fact that the raw observables are strongly suppressed by the noise and for very large $N_2$ tend to cluster around 0. The same DNN can mitigate errors for different initial conditions, the quality of data improvement  remains nearly the same. The quality of error mitigation for very large $N_2$ becomes limited essentially by the statistic uncertainties associated with the probabilistic nature of measurement (shot noise). DNN naturally amplifies the shot noise, since it also amplifies the whole signal. However, the shot noise produces no bias, in contrast to quantum gate errors, which do produce bias. Note that the shot noise on mitigated curves can be additionally smoothed by using some other technique, but we do not address this issue in the present paper.

\subsection{Weight-2 observables}

\begin{figure}
\begin{center}
    \includegraphics[width=0.95\linewidth]{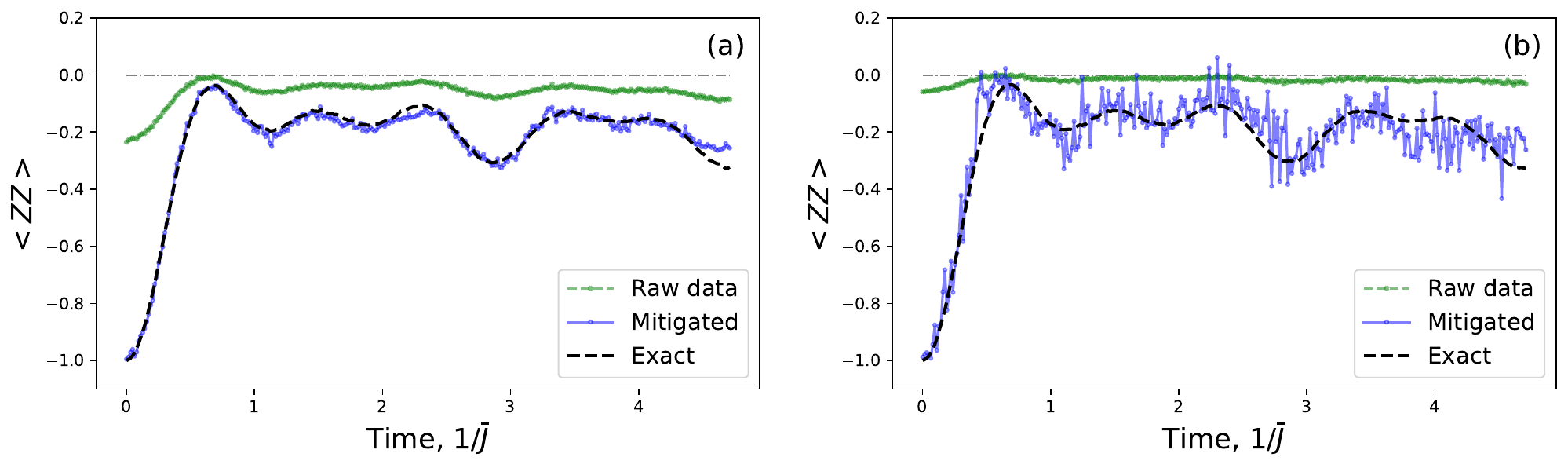}
    \caption{\label{fig:weight2}The dependence of a $\langle ZZ \rangle$ of 9-spin system on time for Trotter layer numbers $N_2=16$ (a), $N_2=32$ (b) starting from the initial condition $\vert 010101010\rangle$ at $\bar{h}=2\bar{J}$. Trotter layer number at the training stage is $N_1=4$. Depolarizing error channel is assumed.}
\end{center}
\end{figure}

In this subsection we concentrate on local weight-2 observables, which, in real experiments, are known \cite{kim2023scalable} to be more sensitive to noise than their weight-1 counterparts. We evaluate $\langle Z_{ij} \rangle$ between nearest neighbors in the spin lattice and address the average local $ZZ$ quantity $\langle ZZ \rangle = \sum_{<ij>}\langle Z_{i}Z_{j} \rangle /N_{adj}$, where $N_{adj}=12$ is the total number of adjacent pairs for 9 qubits. Thus, new DNN is constructed which has 12 input and 12 output neurons. Figure \ref{fig:weight2} shows the dynamics of $\langle ZZ \rangle$ for $N_2=16$ (a) and $N_2=32$ (b) starting from the initial state $\vert 010101010\rangle$ (checkerboard pattern of excited spins). We again see a dramatic improvement of the data quality, which is limited by the statistical uncertainties (shot noise). Note that, according to our simulations, weight-2 observables are indeed more sensitive to noise than their weight-1 counterpart.

\subsection{Scaling}

\begin{figure}
\begin{center}
\includegraphics[width=0.5\linewidth]{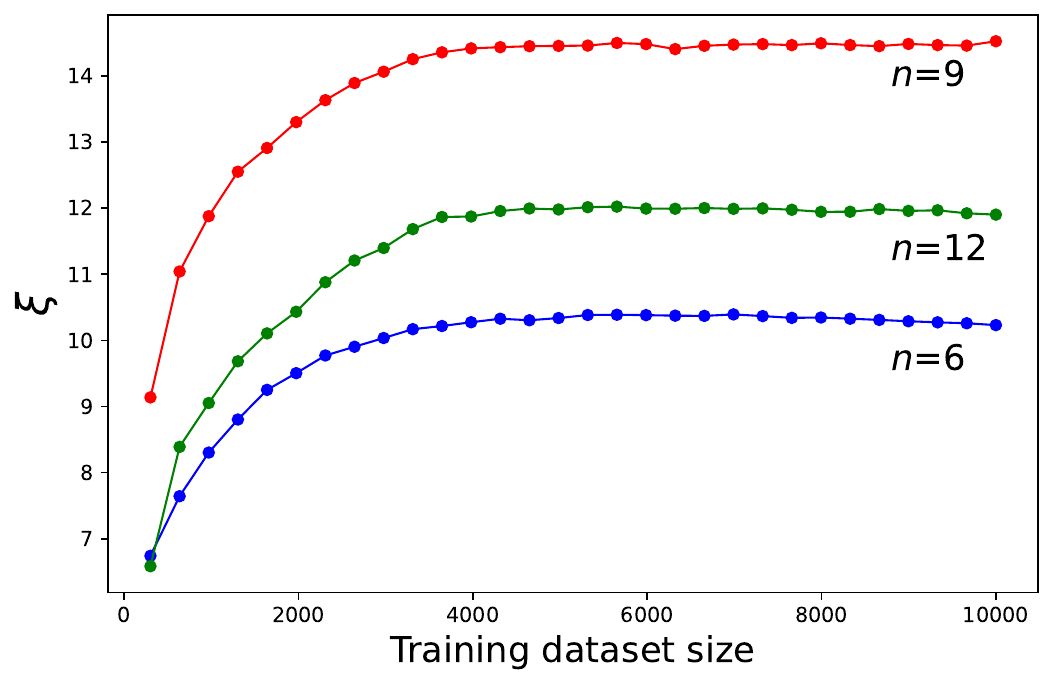}
\caption{\label{fig:train}
The dependence of $\xi$ on training input number for different qubit numbers. Trotter layer numbers are $N_1=4$, $N_2=16$. Depolarizing error channel was assumed.}
\end{center}
\end{figure}

In the present subsection, we address scaling issue. We analyze how quality of error mitigation depends on the number of points randomly chosen in Monte Carlo method. We consider different qubit numbers $n$. For each $n$, we address different number of random inputs in Monte Carlo method and for each number we evaluate mean MSE between the exact and error-mitigated results for the whole set of inputs. The domain of inputs is formed by randomly chosen initial condition (eigenstate in the computational basis) and time instance from the interval $(0,T)$ divided into 100 segments.  We then study the dependence of the MSE on the number of inputs. Namely, we concentrate on the value of $\xi$, which is defined as $\xi=MSE_b/MSE_a$, where $MSE_b$ is MSE before the error mitigation and $MSE_a$ is MSE after the error mitigation. We find smooth dependencies of $\xi$ on input number, which saturate. They are shown in Fig. \ref{fig:train} for $n=6,9, 12$ at $N_1=4$, $N_2=16$. We then see that the typical number of inputs, which leads to the  saturation of $\xi$, is independent on $n$ for such mesoscopic systems despite of the fact that the total number of inputs grows exponentially. 

\subsection{The effect of shot counts}

\begin{figure}
\begin{center}
\includegraphics[width=0.75\linewidth]{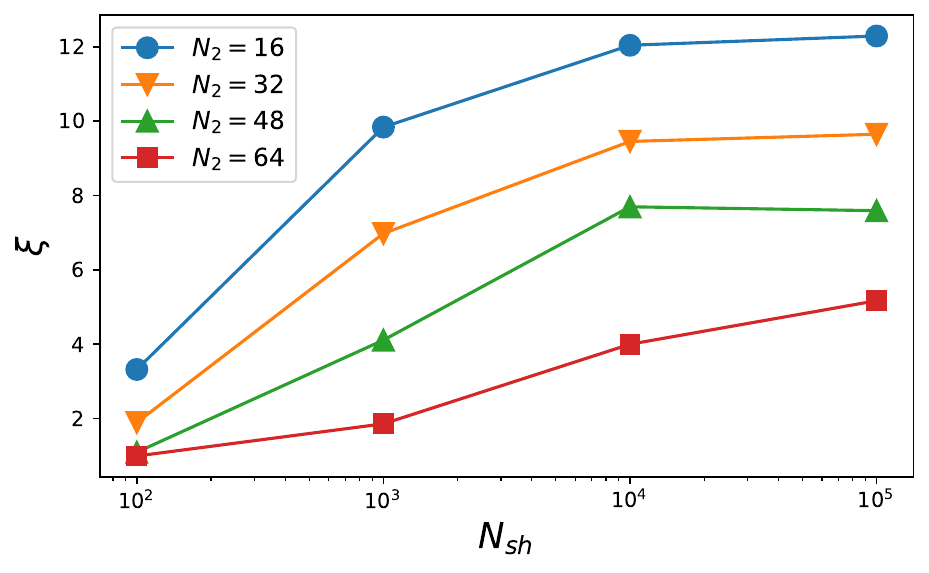}
\caption{The dependence of $\xi$ on shot number $N_{sh}$ for different value of $N_2$ at $N_1=4$. The number of spins in the system $n=6$. Depolarizing error channel was assumed. }
\label{fig:mse}
\end{center}
\end{figure}

We also explored how the efficiency of our method depends on the number of shots (measurements). This is illustrated in Fig. \ref{fig:mse} for $n=6$, where we show $\xi$ as a function of shot number $N_{sh}$ for different values of $N_2$ at $N_1=4$. We see that $\xi$ as a function of $N_{sh}$ generally shows a saturation in the vicinity of $N_{sh} = 10^4$ except of the very large $N_2=64$, when the algorithm tends to break down. Of course, the criterion of the algorithm breakdown, i.e., the characteristic $N_2$ depends also on $p_1$ and $p_2$, so that further improvement of quantum processor quality would increase this number and allow for the implementation of deeper quantum circuits enhanced by DNN.

\section{Non-depolarizing errors}

\subsection{Pauli noise}

Pauli noise is an extension of the depolarizing noise. It is represented by the single-qubit quantum channel
\begin{equation}
    \Phi^{Pauli}_{1q}(\rho) = (1 - p_1)\rho + p_{1x}X\rho X + p_{1y}Y\rho Y + p_{1z}Z\rho Z,
\end{equation}
where $p_1$ is a probability of no error during the single-qubit gate, $p_{1x}$, $p_{1y}$ and $p_{1z}$ are probabilities of occurring an $X$, $Y$, or $Z$ errors, respectively. This noise is reduces to the depolarizing noise provided all these probabilities are the same. For the two-qubit gates the quantum channel is described by Eq. (4).

\begin{figure}
\begin{center}
\includegraphics[width=0.95\linewidth]{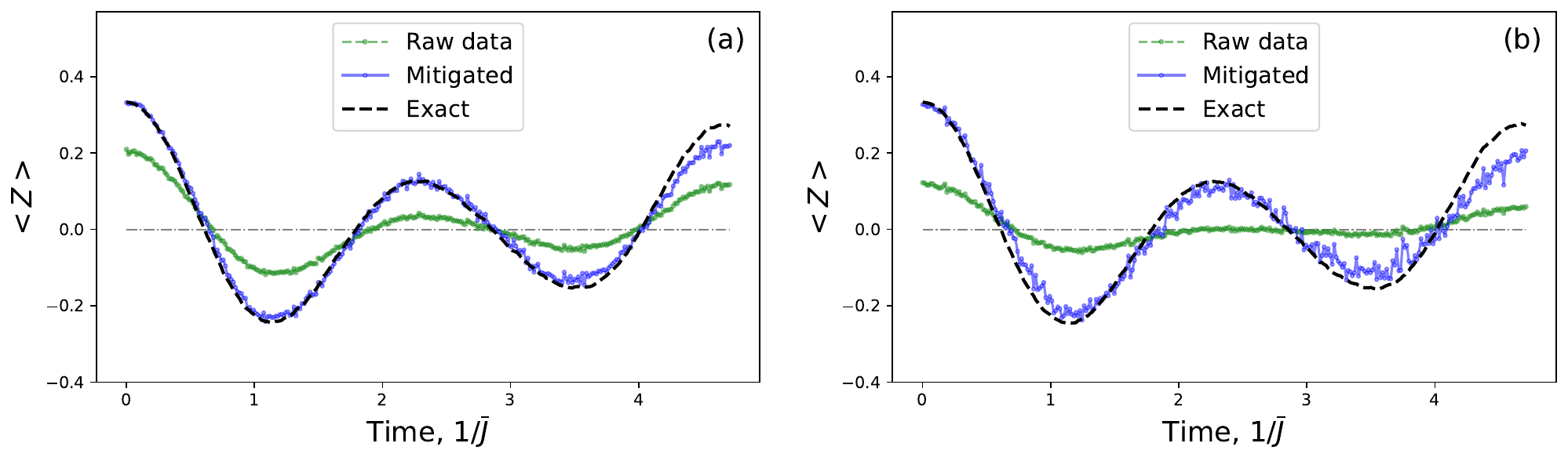}
\caption{\label{fig:weight1p}
The dependence of a mean magnetization in $z$ direction $\langle Z\rangle$ of 9-spin system on time for Trotter layer numbers $N_2=32$ (a) and $N_2=64$ (b) starting from the initial condition $\vert 000111000\rangle$ at $\bar{h}=2\bar{J}$. The inhomogeneous Pauli channel is assumed. Trotter layer number at the training stage is $N_1=4$.}
\end{center}
\end{figure}

\begin{figure}
\begin{center}
\includegraphics[width=0.95\linewidth]{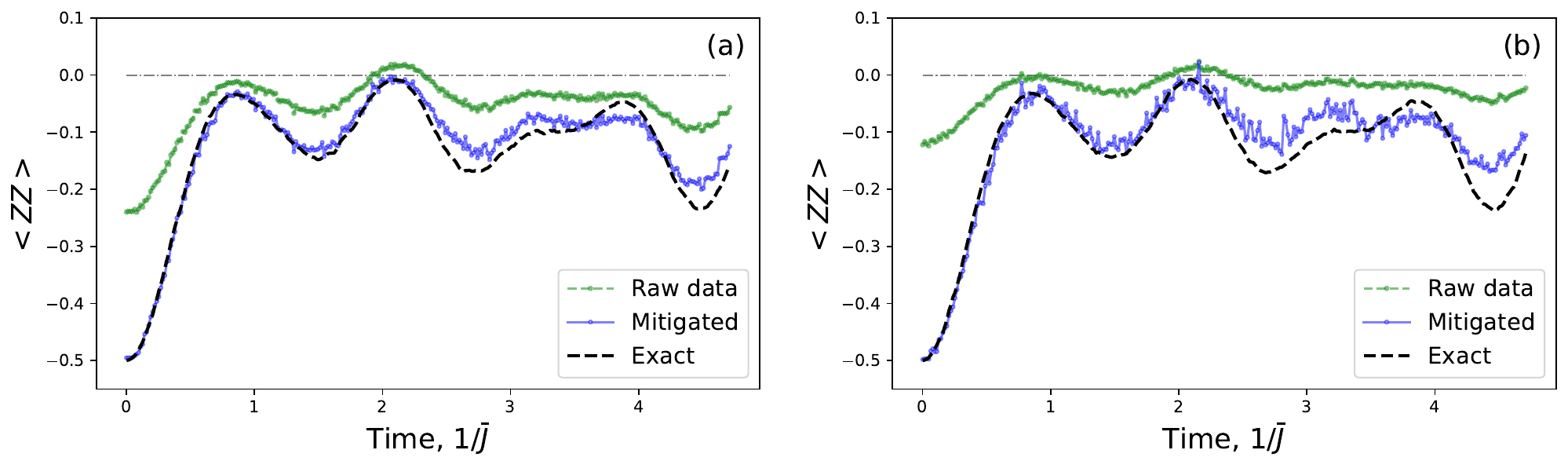}
\caption{\label{fig:weight2p}
The dependence of a $\langle ZZ \rangle$ of 9-spin system on time for Trotter layer numbers $N_2=16$ (a), $N_2=32$ (b) starting from the initial condition $\vert 010101010\rangle$ at $\bar{h}=2\bar{J}$. Trotter layer number at the training stage is $N_1=4$. Inhomogeneous Pauli error channel is assumed.}
\end{center}
\end{figure}

For our numerical experiments, we take $p_{1x}=0.5*10^{-4}$, $p_{1y}=1.0*10^{-4}$, $p_{1z}=2.0*10^{-4}$ and $p_{2x}=1.0*10^{-3}$, $p_{2y}=2.0*10^{-3}$, $p_{2z}=3.0*10^{-3}$. The results for the mean magnetization $\langle Z \rangle$ are shown in Fig. \ref{fig:weight1p} for the initial condition $\vert 000111000\rangle$ at $\bar{h}=2\bar{J}$ and $N_2=32$ (a), $N_2=64$ (b), while $N_1=4$. The quality of error mitigation in this case is similar to that for the depolarizing channel. Figure \ref{fig:weight2p} presents results for weight-2 observable $\langle ZZ \rangle$ for the same set of parameters. The efficiency of error mitigation is high, but it becomes lower than that for the depolarizing channel.

\subsection{Crosstalk noise}

A $ZZ$ crosstalk noise model, relevant for fixed-frequency qubits, is based on $ZZ$ pairwise interaction between qubits for which two-qubit gates can be directly applied. It is represented by the two-qubit quantum channel
\begin{equation}
	\label{eqn:crosstalk dm}
    \Phi(\rho) = U_{ZZ}\rho U_{ZZ}^{\dagger},
\end{equation}
where
\begin{equation}
    U_{ZZ}(t) = 
    \begin{pmatrix}
        e^{-i\zeta t} & 0 & 0 & 0 \\
        0 & e^{i\zeta t} & 0 & 0 \\
        0 & 0 & e^{i\zeta t} & 0 \\
        0 & 0 & 0 & e^{-i\zeta t}
    \end{pmatrix},
\end{equation}
where $\zeta$ is a coupling constant. In our numerical experiments, we used $\zeta = 2\pi*50$ KHz for all pairs of adjacent qubits, which corresponds to typical crosstalks in IBM quantum processors \cite{Babukhin2020}. The time duration of a two-qubit gate in this arcitecture is nearly 400 ns, so that the typical error due to the crosstalk on this scale is of the order of 1 percent. This simple estimate is consistent with the more involved analysis of Ref. \cite{perrin2023mitigating}. 

\begin{figure}
\begin{center}
\includegraphics[width=0.95\linewidth]{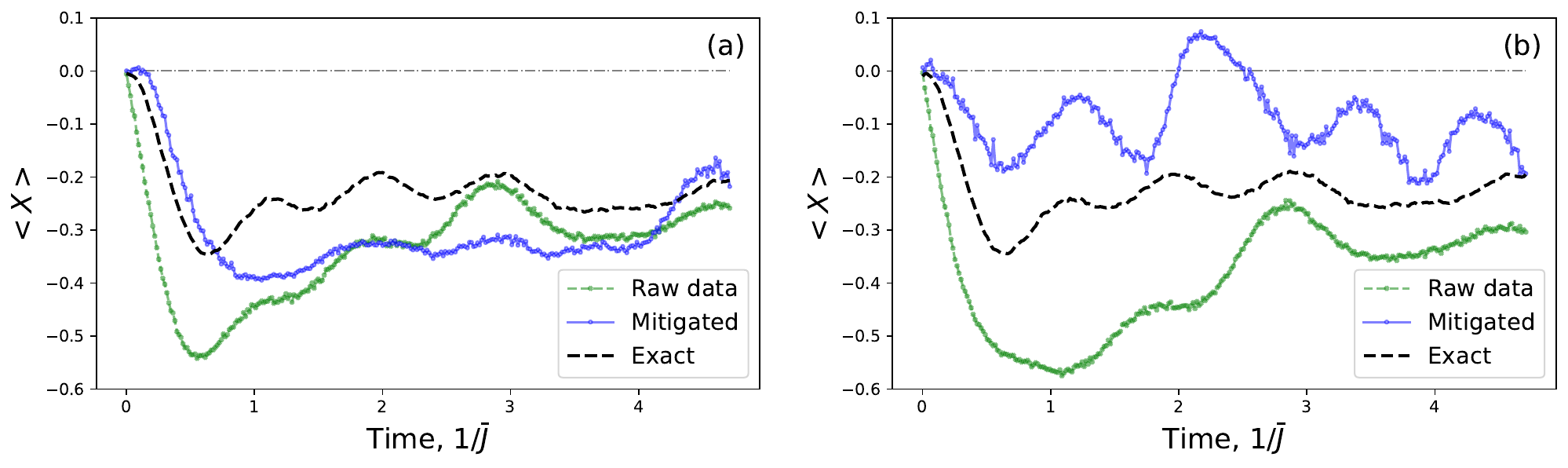}
\caption{\label{fig:crosstalks}
    The dependence of a magnetization in $x$ direction $\langle X \rangle$ of 9-spin system on time for Trotter layer numbers $N_2=16$ (a), $N_2=32$ (b) starting from the initial condition $\vert 010101010\rangle$ at $\bar{h}=2\bar{J}$. Trotter layer number at the training stage is $N_1=4$. Crosstalk noise model is asssumed.}
\end{center}
\end{figure}

The result for the dynamics of magnetization in the $x$ direction $\langle X \rangle = 1/n \sum_j \langle X_j \rangle$ for the initial state $\vert 010101010\rangle$ at $\bar{h}=2\bar{J}$, $N_1=4$ and $N_2=16$ (a), $N_2=32$ (b) is shown in Fig. \ref{fig:crosstalks}. We see that our method fails to improve data quality for the crosstalk noise. The reason can be associated with the complex deformation of curves representing the dependence of magnetization on time induced by crosstalks. We can conclude that the performance of our method for experimental data of Ref. \cite{zhukov2022quantum} was likely limited by $ZZ$ crosstalks. The use of the numerical simulations in the present article allow us to analyze different noises and to understand their effects separately.

\section{Conclusions}

In the present paper, we addressed the impact of different types of quantum noises on a performance of the recently proposed method of quantum error mitigation, which is based on deep neural network application. We considered a Trotterized dynamics of both local weight-1 and weight-2 observables for the 2D spin lattice described by the transverse-field Ising model Hamiltonian. We assume that the target circuit consists of $N_2$ Trotter steps. The general idea of the method is to train a deep neural network to transform noisy data for $N_1<N_2$ Trotter steps into their more exact counterparts. Such data for training can be obtained either by numerical simulation or as outputs from the same quantum computer. In the latter case, the errors can be additionally mitigated by the zero noise extrapolation. In the training circuits, $N_1$ Trotter steps are supplemented by complex identity gates or a single identity gate, which originate from the Trotterized dynamics and which boost the noise rate towards the target circuit with $N_2$ Trotter steps. After that, a trained network is applied to mitigate quantum errors for $N_2$ Trotter steps. In order to obtain an informative training set, different initial conditions for the spin lattice were considered and training circuits were constructed for these initial conditions.

Using numerical simulations, we demonstrated a dramatic improvement of data quality by the DNN for incoherent noises, such as depolarizing and inhomogeneous Pauli channels. The quality of error mitigation for incoherent errors in the regimes of high noise is limited essentially by the statistical errors originating from probabilistic nature of measurements, provided SPAM errors are neglected. However, coherent errors due to $ZZ$ crosstalks are not mitigated. We concluded that such noises should be at first converted into incoherent errors by randomized compiling before the application of our method. 

\section*{Acknowledgements}
Useful discussions with D. V. Babukhin and N. M. Guseynov are acknowledged. 
W. V. P. acknowledges support from the RSF grant No. 23-72-30004 (https://rscf.ru/project/23-72-30004/).

\section*{Data availability}
Raw data for this study were generated by running quantum circuits via QISKIT. The data used in the current study is available upon reasonable request from the corresponding authors. The code used can be found at https://zenodo.org/records/10477667.

\backmatter

\bibliography{main}

\end{document}